\documentclass{appolb}
\usepackage{graphicx}
\usepackage{mathrsfs}
\usepackage{pifont}
\usepackage{yfonts}

\usepackage{graphicx}
\usepackage{dcolumn}
\usepackage{bm}
\usepackage{epsfig}
\usepackage{slashed}
\usepackage[usenames, dvipsnames]{color}

\DeclareMathAlphabet{\mathpzc}{OT1}{pzc}{m}{it}

\definecolor{myoxra}{RGB}{200, 122, 77}
\definecolor{mypink1}{RGB}{177, 77, 122}
\definecolor{mygreen1}{RGB}{0, 111, 0}
\definecolor{myblue1}{RGB}{0, 0, 111}
\definecolor{mygrey1}{RGB}{111, 111,111}
\definecolor{mycolor1}{RGB}{111, 0,0}
\definecolor{mycolor2}{RGB}{111, 111,0}
\definecolor{mycolor3}{RGB}{153, 153,0}
\definecolor{mycyan1}{RGB}{0, 111, 111}
\definecolor{mygold}{RGB}{250, 200, 50}

\newcommand{\beq}{\begin{equation}} 
\newcommand{\eeq}{\end{equation}} 
\newcommand{\bega}{\begin{eqnarray}} 
\newcommand{\ega}{\end{eqnarray}} 

\newcommand{\dhd}{{\textstyle d}
\lower.03ex\hbox{\kern-0.38em$^{\scriptstyle-}$}\kern-0.05em{}}
\newcommand{\dbar}{{\textstyle \delta}
\lower.03ex\hbox{\kern-0.38em$^{\scriptstyle-}$}\kern-0.05em{}}

\newcommand{\barc}{{\bar c}}

\newcommand{\baru}{{\bar u}}

\newcommand{\barv}{{\bar v}}

\newcommand{\Bara}{{\bar A}}

\newcommand{\Barf}{{\bar F}}

\newcommand{\bfi}{{\bar \phi}}
\newcommand{\bsi}{{\bar \psi}}

\newcommand{\barA}{{\bar A}}

\newcommand{\barD}{{\bar D}}
\newcommand{\Bard}{{\bar D}}
 
\newcommand{\barF}{{\bar F}}

\newcommand{\cala}{{\cal A}}

\newcommand{\cald}{{\cal D}}  
\newcommand{\calD}{{\cal D}}
\newcommand{\calF}{{\cal F}}
\newcommand{\calf}{{\cal F}}

\newcommand{\calo}{{\cal O}}

\newcommand{\bala}{\bar{\cal A}}
\newcommand{\balf}{\bar{\cal F}}
\newcommand{\balF}{\bar{\cal F}}

\newcommand{\hatA}{{\hat A}}

\newcommand{\hatF}{{\hat F}}

\newcommand{\hsi}{{\hat \psi}} 
\newcommand{\hbsi}{{\hat {\bar\psi}}}

\newcommand{\tilA}{{\tilde A}}

\newcommand{\tilZ}{{\tilde Z}}
\newcommand{\Tilz}{{\tilde Z}}

\newcommand{\tmu}{\tilde {\mu}}

\newcommand{\ve}{\varepsilon}

\newcommand{\slpar}{\slashed{\partial}}
\newcommand{\slp}{\slashed{p}}
\newcommand{\slq}{\slashed{q}}
\newcommand{\slA}{\slashed{A}}
\newcommand{\slP}{\slashed{P}}

\begin{document}
\title{Background-field method and QCD factorization
\footnote{Submitted to Acta Physica Polonica B special volume dedicated to
Dmitry Diakonov, Victor Petrov and Maxim Polyakov.}
}

%

\author{Ian Balitsky
\address{Physics Dept., Old Dominion University, Norfolk, VA 23529,\\
 and \\
 Theory Group, JLAB, 12000 Jefferson Ave, Newport News, VA 23606}
}

\preprint{JLAB-THY-25-4217}

\maketitle
\begin{abstract}

One method for deriving a factorization for QCD processes is to use successive integration over fields in the functional integral. In this approach, we separate the fields into two categories: dynamical fields with momenta above a relevant cutoff, and background fields with momenta below the cutoff. The dynamical fields are then integrated out in the background of the low-momentum background fields.
This strategy works well at tree level, allowing us to quickly derive QCD factorization formulas at leading order. However, to extend the approach to higher loops, it is necessary to rigorously define the functional integral over dynamical fields in an arbitrary background field. This framework was carefully developed for the calculation of the effective action in a background field at the two-loop level in the classic paper by Abbott \cite{Abbott:1980hw}.
Building on this work, I specify the renormalized background-field Lagrangian and define the notion of the quantum average of an operator in a background field, consistent with the “separation of scales” scheme mentioned earlier. As examples, I discuss the evolution of the twist-2 gluon light-ray operator and the one-loop gluon propagator in a background field near the light cone.

\end{abstract}

\section{Reminiscences}
Unfortunately, I never met Maxim in person; I only knew his works. On the contrary, I was fortunate to spend years with Vitya and Mitya at the LNPI theory group and learned a lot from them. Mitya was always like my "big brother," and Vitya was a peer whom I always looked up to. Let me start with my recollections about Vitya.

I met Vitya in 1970 when we were part of the Leningrad team at the all-Soviet physics Olympiad. A couple of years later, we ended up at Leningrad University in a small group of students studying "elementary particle physics." This was in 1973 when QCD was being formulated, but we only heard about it a couple of years later when Prof. Gribov gave lectures on the theory of strong interactions. 

After graduating from the University, Vitya and I became "aspirants" - graduate students at the Leningrad Nuclear Physics Institute. This period was very exciting; QCD was rapidly developing, and the "Leningrad group" was one of the best places in the world to study QCD from experts like Gribov and Lipatov. Vitya was working on his Ph.D. on a different topic -- Schwinger's model of confinement -- which provided him with a broader understanding of how quantum field theory operates beyond the calculation of Feynman diagrams. I believe that knowledge helped him and Mitya develop an "instanton liquid" model of the QCD vacuum a few years later. 

I must admit that, at first, I did not know much beyond Feynman diagrams, so I often asked Vitya about various problems outside perturbation theory. Needless to say, he was always willing to help. While we were not close friends, we were definitely close peers—we spent 15 years at LNPI in contact. Even after I left for the U.S., we always met whenever I visited St. Petersburg. 

As I mentioned, if Vitya was my peer, Mitya was like a "big brother." During the relatively short time we worked on QCD sum rules in relation to exotic mesons, I learned a great deal from him, not only how to derive formulas but also how to navigate the sometimes uncertain waters of theoretical physics. I remember he said that to become well known, one must study a topic a little bit before the whole community becomes interested in it. If you study this subject five years before or five years later, you waste your time -- nobody will notice. He once joked that he was probably getting older since he no longer jumped to investigate every new idea coming from the outside world. This was a completely new perspective for me, as the culture of the LNPI theoretical department was to study one's subject deeper and deeper, without paying attention to events happening in the outside world. Of course, that approach resulted in having the best experts in perturbative QCD, but "our vices are the continuation of our virtues"—we missed out on some important theoretical developments, supersymmetry being one of them. Mitya was different; after working on pQCD (the famous DDT paper), he switched to non-perturbative physics and, along with Vitya, developed the instanton liquid vacuum model. 

In the years to come, I always remembered Mitya’s imperative: "Always be on alert for new ideas floating around," though I must confess that I did not follow his advice with enough zeal. Needless to say, we kept in touch after I left Leningrad, especially when Mitya spent almost a year working at JLab on pentaquark physics. 

In conclusion, I believe there is no need to praise Mitya's and Vitya’s work here -- for instance, the instanton liquid is now a well-established model of the QCD vacuum. I just want to emphasize that they were not only exceptional physicists but also very good people, and we miss them.

\section{Introduction}
The background-field technique was invented by Schwinger many decades ago and since then 
was extensively used in gauge and gravitational theories, especially for the calculation of the effective action introduced in the papers
\cite{DeWitt:1967uc, Honerkamp:1971sh,tHooft:1973bhk, Arefeva:1974jv}. The effective action approach turned out to be very convenient for the studies of  gauge theories with symmetry breaking in the early years of Standard Model \cite{Coleman:1973jx,Coleman:1974hr}.  
In recent years, the effective action was extensively used for research of the correspondence between  BSM models 
and the so-called SMEFT - low-energy effective field  theory studying possible effects of addition of higher-dimension operators 
to SM Lagrangian, see e.g. Ref. \cite{Fuentes-Martin:2024agf}.

In contrast, in QCD, the background-field method has been applied beyond the effective action, primarily to derive factorization formulas for QCD processes by using successive integration over fields in the functional integral.  The classical example is the QCD sum rules \cite{Shifman:1978bx}, where at first we integrate over quark and gluon fields with hard momenta  and get perturbative diagrams for coefficient functions in front of the local operators with sort momenta  (vacuum condensates). At this step, it is convenient to treat soft fields  as background fields
 and use the background-field method. Technical aspects of using the background field approach for QCD sum rules were discussed in Ref. \cite{Novikov:1984ecy}.

 Another application of the background-field method is the study of deep inelastic scattering (DIS) using the light-cone expansion in light-ray operators \cite{Balitsky:1987k}. Similar to the approach in QCD sum rules, we begin by integrating over quark and gluon fields with transverse momenta $k_\perp$ greater than some factorization scale $\mu$. This step yields coefficient functions that multiply the light-ray operators with transverse momenta up to $\mu$.
 At this stage,  it is again convenient to treat fields with $k_\perp<\mu$  as background fields and use 
 the expansion of propagators in the background-field gauge near the light cone \cite{Balitsky:1987k}. 
 Recently, this technique has been employed to derive matching relations between lattice calculations of gluon pseudo-PDFs and conventional light-cone gluon PDFs \cite{Balitsky:2019krf, Balitsky:2021qsr, Balitsky:2021cwr}.

The background-field method was also used to study the  rapidity factorization for high-energy scattering. To understand the high-energy behavior of a QCD amplitude, one integrates over fields with rapidity greater than some “rapidity divide” $\eta$, yielding impact factors - coefficient functions multiplying Wilson-line operators 
with rapidity smaller than $\eta$ \cite{Balitsky:1995ub}.   In this case, for the purpose of calculating impact factors, gluons (and quarks) with rapidity $Y<\eta$  are treated as background fields. More recently, the background-field technique has been applied to derive evolution equations and power corrections to TMD factorization
 \cite{Balitsky:2015qba,Balitsky:2020jzt,Balitsky:2024ozy,Vladimirov:2023aot,Scimemi:2019gge}.

Remarkably,  in all these cases the naive picture of separation between dynamical fields with momenta above some cutoff and background
fields with momenta below the cutoff works pretty well at the tree level enabling us to quickly get QCD factorization formulas 
at the leading order. However, to get to higher loops one needs to rigorously define the functional integral over dynamical fields in an arbitrary background field.

This program was carefully implemented for the calculation of the effective action in a background field at the two-lop level 
in the classical paper \cite{Abbott:1980hw}. As demonstrated there, if one uses background-field gauge which preserves the gauge invariance, one can renormalize only background fields and leave the quantum dynamical fields inside the loops unrenormalized.

However, to go beyond the effective action and to calculate, for example, the light-cone expansion of one-loop propagator in the background field one needs to take into account also the renormalization of quantum fields. In this  paper
I specify the renormalized Lagrangian in the background field and define the notion of quantum average of an operator
in the background field consistent with naive ``separation of scales'' scheme. As examples, I consider the evolution
of twist-2 gluon light-ray operator and one-loop gluon propagator near the light cone.

The paper is organized as follows. In Sect. 2 I define the renormalized Lagrangian in the gluon background field and in Sect. 3 I illustrate diagrams which ensure the requirement that 
a single quantum field cannot turn to background field(s) in accordance with naive factorization setup. The  evolution
of twist-2 gluon light-ray operator is discussed in Sect. 4  and one-loop gluon propagator near the light cone in Sect. 5. 
The Appendices contain necessary technical details.

\section{Renormalized Lagrangian in a background field}
First, let us briefly remind how to get an effective action in the background-Feynman gauge following Abbott's approach \cite{Abbott:1980hw,Abbott:1981ke}. One defines the  generating functional in the background-Feynman (bF) gauge by the expression

\beq
Z(J,\barA)~=~e^{iW(J,\Bara)}~=~\int\!\calD A_\mu\calD\barc\cald c\cald\bsi\cald \psi~e^{i\!\int\! dz [L(A+\barA)+\half (\Bard_\mu A^a_\mu)^2-A_\mu J^\mu ]}
\label{Ze}
\eeq
where $\Bara$ is the background field,   $L$ is the QCD Lagrangian including ghosts corresponding to  bF gauge-fixing term $\half (\Bard_\mu A^a_\mu)^2$). We use standard notation $\Bard_\mu A^a_\nu~\equiv~\partial_\mu A^a_\nu +gf^{abc}\barA_\mu^b A_\nu^c$
for the the covariant derivative. 

The effective action is defined as a Legendre transform of $W(J,\Bara)$
\beq
\Gamma(\tilA,\barA)~=~W(J_\tilA,\Bara)-\int\! dx J_{\tilA,\barA}^{a,\mu}\tilA^a_\mu
\label{effact}
\eeq
where the source $J_{\tilA,\barA}$ is that  which produces field $\tilA$
\beq
\tilA_\mu^a(x)~=~{\delta W(J,\barA)\over\delta J_\mu^a(x)}~=~
\int\!\calD\Phi A_\mu(x)~e^{i\!\int\! dz \big[L(A+\barA)+\half (\Bard_\mu A^a_\mu)^2-A_\mu J^\mu_{\tilA,\barA}\big]}
\label{tila}
\eeq
Hereafter we denote $\calD A_\mu\calD\barc\cald c\cald\bsi\cald \psi$ by $\cald\Phi$ for brevity.
As demonstrated in Ref. \cite{Abbott:1980hw}, the effective action $\Gamma(0,\barA)$ 
defined as 
\beq
\Gamma(\barA)~\equiv ~\Gamma(0,\barA)~=~W(J_0,\Bara)
\label{effaction}
\eeq
with a source $J_{0,\barA}=J_0(\barA)$ producing zero field $\tilA$
\beq
0~=~
\int\!\calD\Phi A_\mu(x)~e^{i\!\int\! dz [L(A+\barA)+\half (\Bard_\mu A^a_\mu)^2-A^a_\mu J^{a,\mu}_0(\barA)]},
\label{tilazero}
\eeq
describes a sum of one particle irreducible (1PI) diagrams with $\barA$ fields as external legs and quantum $A$ fields inside loops. 

Similarly to Eq. (\ref{effaction}), I define vacuum average of operator $\calo$ in the background field $\Bara$
in the background-Feynman gauge by the formula
\footnote{For simplicity, we do not consider background quark or ghost fields}
\bega
&&\hspace{-1mm}
\langle\hat\calo(A+\barA)\rangle_\barA~
\label{defiObF}\\
&&\hspace{-1mm}=~
{\int\!\calD \Phi~\calo(A+\barA,\psi)~e^{i\!\int\! dz \big[ L(A+\barA,\psi,c)+\half (\Bard_\mu A^a_\mu)^2-A^a_\mu J_0^{a,\mu}(\barA)\big] }
\over \int\!\calD\Phi~e^{i\!\int\! dz\big[ L(A+\barA,\psi,c)+\half (\Bard_\mu A^a_\mu)^2- A^a_\mu J_0^{a,\mu}(\barA)\big] }}
\nonumber
\ega
The linear term $A^a_\mu J_0^{a,\mu}(\barA)$ in the exponent yields
\beq
\langle \hatA\rangle_\barA~=~0,
\label{netua}
\eeq
 see Eq. (\ref{tilazero}). As in the case of effective action (\ref{effaction}), this equation  ensures that there is no transition of quantum field $A$ to the background field $\barA$. 
 This property is in accordance with naive factorization requirement that the  ``dynamical'' field
 with relevant component of the momentum above some cutoff cannot go to the ``background'' field(s) with the momentum 
 below the cutoff.

In the leading order $J_0^{a,\mu}=\Bard_\alpha \barF^{a,\alpha\mu}$ so the
exponent in Eq. (\ref{defiObF}) takes the form (for $n_f$ flavors of  massless quarks)
\begin{eqnarray}
&&\hspace{-0mm} 
L(A+\Bara)+\half (\Bard_\mu A^a_\mu)^2-A_\nu\Bard_\mu \Barf^{\mu\nu}~
\nonumber\\  
&&\hspace{-1mm} 
=~-{1\over 4}\big(\Barf^a_{\mu\nu}\big)^2
+\half A^{a\mu}(\Bard^2 g_{\mu\nu}-2ig\barF_{\mu\nu})^{ab}A^{b\nu}
-\barc^a\Bard^\mu(\Bard_\mu -ig\mu^\ve A_\mu)^{ab}c^b
\nonumber\\  
&&\hspace{-1mm} 
+~\sum_f\bsi i\slashed{\barD}\psi
-~g f^{abc}\Bard_\mu A^a_\nu A^b_\mu A^c_\nu-{g^2\over 4}(f^{abc}A^b_\mu A^c_\nu)^2
+g\sum_f\bsi\slashed{\barA}\psi
\end{eqnarray}
which leads to ``bare'' propagators in background fields
\begin{eqnarray}
&&\hspace{-1mm} 
\langle \hatA^a_\mu(x) \hatA^b_\nu(y)\rangle_\barA~=~
\int\!\calD\Phi~A^a_\mu(x) A^b_\nu(y)~e^{i\int\! dz[L(A+\barA)+\half (\Bard_\mu A^a_\mu)^2-A_\nu\Bard_\mu \Barf^{\mu\nu}] }
\nonumber\\  
&&\hspace{11mm} 
=~(x|{-i\over P^2+2ig\barF+i\epsilon}|y)^{ab}_{\mu\nu},
\nonumber\\  
&&\hspace{-1mm} 
\langle \hat{c}^a(x) \hat\barc^b(y)\rangle_\barA~=~
\int\!\calD \Phi~c^a(x) \barc^b(y)~e^{i\int\! dz[L(A+\barA)+\half (\Bard_\mu A^a_\mu)^2-A_\nu\Bard_\mu \Barf^{\mu\nu}] }
\nonumber\\  
&&\hspace{11mm} 
=~(x|{i\over P^2+i\epsilon}|y)^{ab},
\label{props}\\  
&&\hspace{-1mm} 
\langle \hsi(x) \hsi(y)\rangle_\barA~=~
\int\!\calD \Phi~\psi(x) \bsi(y)~e^{i\int\! dz[L(A+\barA)+\half (\Bard_\mu A^a_\mu)^2-A_\nu\Bard_\mu \Barf^{\mu\nu}] }
\nonumber\\  
&&\hspace{11mm} 
=~(x|{i\over \slashed{P}+i\epsilon}|y)~\equiv~(x|\slashed{P}{i\over P^2+\half\sigma F+i\epsilon}|y)
\nonumber
\end{eqnarray}
where $\sigma F\equiv \sigma_{\xi\eta} F^{\xi\eta}=\half[\gamma_\xi,\gamma_\eta]F^{\xi\eta}$.
Here we use Schwinger's notations 
$(x|y)=\delta(x-y),~(x|p_\mu|y)=i{\partial\over\partial x^\mu}\delta^4(x-y)$, and $(x|A_\mu|y)=A_\mu(x)\delta(x-y)$.
The operator $P_\mu$ is defined by $P_\mu^{ab}=iD_\mu^{ab}=i\partial_\mu\delta^{ab}-igf^{abc}\barA_\mu^c$ and 
the RHS of the first equation (\ref{props}) is understood as
\bega
&&\hspace{-1mm}
(x|{1\over P^2+2ig\barF}|y)_{\mu\nu}~\equiv~(x|{g_{\mu\nu}\over P^2}
\\
&&\hspace{-1mm}
-~2ig{1\over P^2}\barF_{\mu\nu}i{1\over P^2}
-4g^2{1\over P^2}\barF_{\mu\xi}{1\over P^2}\barF^\xi_{~\nu}
+8ig^3{1\over P^2}\barF_{\mu\xi}{1\over P^2}\barF^{\xi\eta}
{1\over P^2}\barF_{\eta\nu}+...|y)
\nonumber
\ega

Next, let us discuss renormalization. Without external field, the $\overline{MS}$-renormalized QCD Lagrangian in Feynman gauge has the form
\bega
&&\hspace{-1mm}
L^F_{\rm ren}~
=~-{1\over 4}Z_3\Big[ \partial_\mu A^a_\nu-\mu\leftrightarrow\nu+g\mu^\ve f^{abc}{Z_1\over Z_3}A^b_\mu A^c_\nu\Big]^2-{(\partial_\mu A^a_\mu)^2\over 2}
\nonumber\\
&&\hspace{-1mm} 
-~\Tilz_3\barc^a\partial^\mu(\partial_\mu -ig\mu^\ve {\tilZ_1\over\tilZ_3}A_\mu)^{ab}c^b
+Z_2\sum_f\bsi\Big(i\slpar+g\mu^\ve{Z_1^F\over Z_2}\slA\Big)\psi
\label{RenLag}
\ega
wher $\mu$ is the normalization point and $\ve=2-{d\over 2}$. Note that ${Z_1\over Z_3}={\tilZ_1\over\tilZ_3}={Z_1^F\over Z_2}$ due to Ward identities, see e.g. the textbook \cite{Weinberg:1996kr}.

In the absence of a background field, the counterterms $Z_i$ regularize all UV divergencies in Feynman diagrams.
However, in the case of the background field these counterterms are not sufficient to make all Green functions  
(\ref{defiObF}) UV finite. 
For example, let us  consider v.e.v. of integral  (\ref{defiObF}) with $\calo=1$. In the first order in $g^2$ one obtains 
(see e.g. the textbook \cite{Weinberg:1996kr}).
\bega
&&\hspace{-1mm}
\int\!\calD\Phi~e^{iS(A+\barA)+\half (\Bard_\mu A^a_\mu)^2-{\rm source~ term} }
\nonumber\\
&&\hspace{-1mm} 
\simeq~\int\!\calD \Phi~e^{i\!\int\! \big[dz\half A^{a\mu}(\Bard^2 g_{\mu\nu}-ig\barF_{\mu\nu})^{ab}A^{b\nu}+\barc^a \Bard^2_{ab}c^b
+\bsi(i\slpar +g\slashed{\barA})\psi}\big]
\label{det}\\
&&\hspace{-1mm} 
=~\exp\Big\{{ig^2b_0\over (4\pi)^2\ve}\!\int\! dx~\barF^a_{\lambda\rho}(x)\barF^{a,\lambda\rho}(x)+{\rm UV~finite~terms}\Big\}
\nonumber
\ega
where $b_0={11\over 12}N_c-{2\over 3}n_f$
Thus, to ensure UV finiteness of Eq. (\ref{det}) we need to add counterterm 
$-{ig^2b_0\over (4\pi)^2\ve}\!\int\! dx~\barF^a_{\lambda\rho}(x)\barF^{a,\lambda\rho}(x)$ to the Lagrangian,
which means renormalization $\barA^{(0)}=(1+ \half\delta Z)\barA$ with $\delta Z= {g^2b_0\over 16\pi^2\ve}$.
In general, the background field $\Bara_\mu$ is renormalized by the factor $Z=Z_3^3Z_1^{-2}$ such that  
$\barA_0=Z^\half\barA$
$$
g_0\barA_\mu^{(0)}~=~g\mu^\ve\barA
$$
so that the covariant derivative $\barD_\mu=\partial_\mu-ig\mu^\ve \barA_\mu$ remains gauge-invariant after renormalization 
(recall that $g_0=g(\mu)\mu^\ve Z^{-\half}$). 
For this reason, it is convenient to define  
\beq
\bala^a_\mu~\equiv ~g\mu^\ve\barA^a_\mu,~~~~~~\balf^a_{\mu\nu}~\equiv~ g\mu^\ve\barF^a_{\mu\nu}~=~\partial_\mu\cala^a_\nu-\partial_\nu\cala^a_\mu+f^{abc}\cala^b_\mu\cala^c_\nu
\eeq
so that the background field 
$\cala_\mu$ does not depend on the renormalization point and may be chosen, for example, as $e^\lambda_\mu(k) e^{ikx}$ to compare with conventional calculations 
involving matrix elements between gluon states.
Also, note that $[\barD_\mu,\barD_\nu]=-i\balf_{\mu\nu}$.

Thus, the renormalized Lagrangian in the background field $\barA$ has the form
\begin{eqnarray}
&&\hspace{-1mm} 
L(A+\barA,c,\psi)-{\rm linear ~source~term}~
\label{Lagrangian}\\  
&&\hspace{-1mm} 
=-{1\over 4}\Big[ Z^\half\Barf^a_{\mu\nu}+Z_3^\half(\Bard_\mu A^a_\nu-\mu\leftrightarrow\nu)
+g\mu^\ve f^{abc}{Z_1\over Z_3^{1/2}}A^b_\mu A^c_\nu)\Big]^2-\half (\Bard_\mu A^a_\mu)^2
\nonumber\\  
&&\hspace{-1mm} 
-~
\Tilz_3\barc^a\Bard^\mu(\Bard_\mu -ig\mu^\ve{\tilZ_1\over\tilZ_3}A_\mu)^{ab}c^b
+Z_2\sum_f\bsi(i\slashed{\barD}+g\mu^\ve{Z_1^F\over Z_2}\slA)\psi -A^a_\nu J_0^\nu(\barA)
\nonumber\\  
&&\hspace{-1mm} 
=~-{Z\over 4}(\Barf^a_{\mu\nu})^2
+{Z_3\over 2} A^{a\mu}(\Bard^2 g_{\mu\nu}-2i\balf_{\mu\nu})^{ab}A^{b\nu}
-~Z_1g\mu^\ve f^{abc}\Bard_\mu A^a_\nu A^b_\mu A^c_\nu
\nonumber\\  
&&\hspace{-1mm} 
-~{g^2\mu^{2\ve}\over 4}Z_1^2Z_3^{-1}(f^{abc}A^b_\mu A^c_\nu)^2
-\Tilz_3\barc^a\Bard^\mu(\Bard_\mu -ig\mu^\ve{\tilZ_1\over\tilZ_3}A_\mu)^{ab}c^b
+{Z_3-1\over 2}
\nonumber\\  
&&\hspace{-1mm} 
\times~(\Bard^\mu A^a_{\mu})^2
+~Z_2\sum_f\bsi\big(i\slashed{\barD}
+g\mu^\ve{Z_1^F\over Z_2}\slA\big)\psi +A^a_\nu
\big[ Z^\half Z_3^\half\barD_\mu \barF^{a,\mu\nu}-J_0^{a,\nu}(\Bara)\big]
\nonumber
\ega
With one-loop accuracy it can be rewritten as
\bega
&&\hspace{-1mm} 
L(A+\barA,c,\psi)-A^a_\nu J_0^{a,\nu}(\Bara)
\nonumber\\  
&&\hspace{-1mm} 
=~-{1\over 4}(\Barf^a_{\mu\nu})^2
+\half A^{a\mu}(\Bard^2 g_{\mu\nu}-2i\balf_{\mu\nu})^{ab}A^{b\nu}
-\barc^a\Bard^\mu(\Bard_\mu -ig\mu^\ve A_\mu)^{ab}c^b
\nonumber\\  
&&\hspace{-1mm} 
+~\sum_f\bsi i\slashed{\barD}\psi
-~g\mu^\ve f^{abc}\Bard_\mu A^a_\nu A^b_\mu A^c_\nu-~{g^2\mu^{2\ve}\over 4}(f^{abc}A^b_\mu A^c_\nu)^2
+g\mu^\ve\bsi\slA\psi
\nonumber\\  
&&\hspace{-1mm} 
-~{1\over 4}\delta Z(\Barf^a_{\mu\nu})^2
+\half \delta Z_3A^{a\mu}(\Bard^2 g_{\mu\nu}-2i\balf_{\mu\nu}-\barD_\mu\barD_\nu)^{ab}A^{b\nu}
\nonumber\\  
&&\hspace{-1mm} 
-~\barc^a\Bard^\mu(\delta\Tilz_3\Bard_\mu -ig\mu^\ve\delta\Tilz_1A_\mu)^{ab}c^b
\label{Lagrangian_bF}\\
&&\hspace{-1mm} 
-~\delta Z_1g\mu^\ve f^{abc}\Bard_\mu A^a_\nu A^b_\mu A^c_\nu
-{g^2\mu^{2\ve}\over 4}(2\delta Z_1-\delta Z_3)(f^{abc}A^b_\mu A^c_\nu)^2
+~\delta Z_2\sum_f\bsi i\slashed{\barD}\psi
\nonumber\\  
&&\hspace{-1mm} 
+~\delta Z_1^Fg\mu^\ve\sum_f\bsi\slA\psi
+~A^a_\nu\big[\half(\delta Z+\delta Z_3)\barD_\mu \barF^{a,\mu\nu}-\delta J_0^{a,\nu}\big]
\nonumber
\end{eqnarray}
where
\begin{eqnarray}
&&\hspace{-1mm} 
\delta Z_3~=~{g^2\over 16\pi^2\ve}\Big({5\over 3}N_c-{2\over 3}n_f\Big),~~~~~~
\delta Z_1~=~{g^2\over 16\pi^2\ve}\Big({2\over 3}N_c-{2\over 3}n_f\Big),~~~~~~~~
\nonumber\\  
&&\hspace{-1mm} 
\delta Z_2~=~-{g^2\over 16\pi^2\ve}{N_c^2-1\over 2N_c},~~~~~~~~~~
\delta Z~=~{g^2\over 16\pi^2\ve}\Big({11\over 3}N_c-{2\over 3}n_f\Big),
\label{Zisoneloop}\\ 
&&\hspace{-1mm} 
\delta\tilZ_3~=~-\delta\tilZ_1~=~{g^2\over 16\pi^2\ve}{N_c\over 2},~~~~~
\delta Z_1^F~=~ -{g^2\over 16\pi^2\ve}\Big(N_c+{N_c^2-1\over 2N_c}\Big)
\nonumber
\end{eqnarray}
Also, in this order
\beq
\delta J_0^\nu(x)~=~{g^2N_c\over 16\pi^2}\Big({8\over 3}N_c-{2\over 3}n_f\Big)(x|\ln{\mu^2\over p^2}|z)
\big[\partial^2 \Bara_\mu(z)-\partial^\nu\partial_\mu \barA_\nu(z)\big]
\label{deltage1}
\eeq
as we will demonstrate below.

\section{First perturbative diagrams for the source}
It is instructive to see how the linear term in Eq. (\ref{Lagrangian_bF}) (the last line in the RHS) ensures the condition $\langle A_\mu\rangle_\Bara=0$. 
\begin{figure}[htb]
\includegraphics[width=125mm]{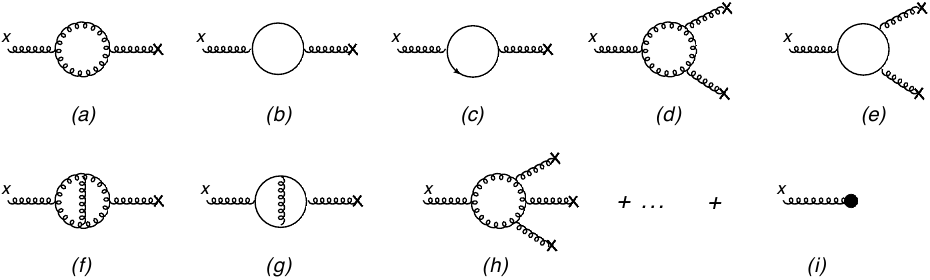}
\caption{Quantum field in $\barA$ background. Tails with crosses denote background fields, black circle denotes source}
\label{fig:bffig1}
\end{figure}
At the $g^2$ level, only the first three diagrams in Fig. \ref{fig:bffig1} contribute. The result of the calculation of the gluon loop in the diagram in Fig.  \ref{fig:bffig1}a is
\begin{eqnarray}
&&\hspace{-1mm}
\int\! {\dhd p\over 2i} ~{1\over (p^2+i\epsilon)[(q-p)^2+i\epsilon]}\Gamma_{\mu\nu,\lambda}(p,q-p)\Gamma^{\rm bF}_{\mu\nu,\rho}(p,q-p)~
\nonumber\\  
&&\hspace{-1mm} 
=~{\Gamma\big(2-{d\over 2}\big)\over (4\pi)^{d\over 2}(-q^2)^{2-{d\over 2}}}B\big({d\over 2},{d\over 2}-1\big)
5(q^2 g_{\lambda\rho}-q_\lambda q_\rho)
\label{pertvkladglu}
\end{eqnarray}
where
\begin{eqnarray}
&&\hspace{-1mm}
\Gamma_{\mu\nu,\lambda}(p,q-p)~=~(2p-q)_\rho g_{\mu\nu}+(-q-p)_\nu g_{\mu\rho}+(2q-p)_\mu g_{\nu\rho}
\label{vertex}
\end{eqnarray}
is (proportional to) the tree-gluon vertex while
\begin{eqnarray}
&&\hspace{-1mm}
\Gamma^{\rm bF}_{\mu\nu,\lambda}(p,q-p)~=~(2p-q)_\rho g_{\mu\nu}+2(q^\mu g^{\nu\gamma}-q^\nu g^{\mu\gamma})
\label{vertexbF}
\end{eqnarray}
is a similar vertex for the emission of a background field $\Bara$ by two quantum gluons which can be read off the term 
$\half A^{a\mu}(\Bard^2 g_{\mu\nu}-2i\balf_{\mu\nu})^{ab}A^{b\nu}$ in the Lagrangian (\ref{Lagrangian_bF}). (The full list of 
Feynman rules in the background-Feynman gauge is presented in Ref. \cite{Abbott:1980hw}).

The ghost contribution in Fig.  \ref{fig:bffig1}b is proportional to
\begin{eqnarray}
&&\hspace{-1mm}
-\int\! {\dhd p\over i} ~{p_\gamma(2p-q)_\rho\over (p^2+i\epsilon)[(q-p)^2+i\epsilon]}~=
~{\Gamma\big(2-{d\over 2}\big)\over (4\pi)^{d\over 2}(-q^2)^{2-{d\over 2}}}B\big({d\over 2},{d\over 2}-1\big)
{q^2 g_{\gamma\rho}-q_\gamma q_\rho\over d-1}
\nonumber\\
\end{eqnarray}
and the quark one  in Fig.  \ref{fig:bffig1}c to
\begin{eqnarray}
&&\hspace{-1mm}
-\int\! {\dhd p\over i} ~{{\rm Tr}\{\slp\gamma_\lambda(\slp-\slq)\gamma_\rho\}\over (p^2+i\epsilon)[(q-p)^2+i\epsilon]}
~=~-8(q^2 g_{\lambda\rho}-q_\lambda q_\rho)B\big({d\over 2},{d\over 2}\big){\Gamma\big(2-{d\over 2}\big)\over (4\pi)^{d\over 2}(-q^2)^{2-{d\over 2}}}
\nonumber\\
\end{eqnarray}
The sum of these contributions with corresponding color and flavor factors is
\begin{eqnarray}
&&\hspace{-1mm}
(q^2 g_{\lambda\rho}-q_\lambda q_\rho){B\big({d\over 2},{d\over 2}-1\big)\Gamma\big(2-{d\over 2}\big)\over (4\pi)^{d\over 2}(-q^2)^{2-{d\over 2}}}\Big[{5d-4\over d-1}N_c-{2(d-2)\over(d-1)}n_f\Big]
\nonumber\\  
&&\hspace{-1mm} 
=~(q^2 g_{\lambda\rho}-q_\lambda q_\rho)\Big[{1\over\ve}+\ln{\mu^2\over q^2}\Big]\Big({8\over 3}N_c-{2\over 3}n_f\Big)
~+~O(\ve)
\nonumber\\
\end{eqnarray}
so we get
\begin{eqnarray}
&&\hspace{-1mm}
\langle \hatA_\mu(q)\rangle_\Bara ~\stackrel{\rm Fig.~\ref{fig:bffig1} a-c}=~{g^2\over 16\pi^2}
{q_\lambda q_\rho-q^2 g_{\lambda\rho}\over q^2}\Big[{1\over\ve}+\ln{\mu^2\over q^2}\Big]
\Big({8\over 3}N_c-{2\over 3}n_f\Big)\barA(q)
\nonumber\\  
&&\hspace{-1mm} 
=~{g^2\over 16\pi^2q^2}\Big({8\over 3}N_c-{2\over 3}n_f\Big)\Big[{1\over\ve}+\ln{\mu^2\over q^2}\Big]
\big[\partial^2 \Bara_\mu(q)-\partial^\nu\partial_\mu A_\nu(q)\big]
\nonumber\\
\end{eqnarray}
which corresponds to 
\begin{eqnarray}
&&\hspace{-3mm}
\langle \hatA_\mu(x)\rangle_\Bara ~
\label{aotvet1}\\
&&\hspace{-3mm}\stackrel{\rm Fig.~\ref{fig:bffig1} a-c}=
{g^2\over 16\pi^2}\Big({8\over 3}N_c-{2\over 3}n_f\Big)\!\int\! dz(x|{1\over p^2}\big[{1\over\ve}+\ln{\mu^2\over p^2}\big]|z)
\big[\partial^2 \Bara_\mu(z)-\partial^\nu\partial_\mu A_\nu(z)\big]
\nonumber
\end{eqnarray}

Now, to get full expression for $\langle A_\mu(x)\rangle_\Bara$ in the $g^2$ order we need to take into account the contribution of the linear term in Eq. (\ref{Lagrangian_bF})
\begin{eqnarray}
&&\hspace{-1mm}
\langle \hatA_\mu(x)\rangle_\Bara 
~\stackrel{\rm Fig.~\ref{fig:bffig1}i}=~{\delta Z+\delta Z_3\over 2}(x|{1\over p^2}|z)\barD^\xi \barF_{\xi\mu}(z)
\label{aotvet2}\\  
&&\hspace{-1mm} 
-~{g^2\over 16\pi^2}\Big({8\over 3}N_c-{2\over 3}n_f\Big)\!\int\! dz(x|{1\over p^2}\big[{1\over\ve}+\ln{\mu^2\over p^2}\big]|z)
\big[\partial^2 \Bara_\mu(z)-\partial^\nu\partial_\mu A_\nu(z)\big]
\nonumber
\end{eqnarray}
It is easy to see that ${\delta Z+\delta Z_3\over 2}$ term exactly cancels the UV contribution in Eq. (\ref{aotvet1}). As to the finite
 term in Eq. (\ref{aotvet1}), the finite source $\delta J$ was chosen to exactly cancel it. 
 Thus, with the definition (\ref{defiObF})
 we get $\langle A\rangle_\barA=0$ at the $g^2$ level in the first order in the background field.  As demonstrated in the Appendix, 
 the second line in Eq. (\ref{aotvet2}) is  generalized to a gauge-invariant expression if all orders in background field are taken into account
 at the $g^2$ level,  and after that the first line in Eq. (\ref{aotvet2}) is canceled exactly. 
 In higher orders in $g$ the source should be chosen in a way to ensure $\langle A\rangle_\barA=0$ property at every order. Note, however, that by gauge invariance the only UV-divergent linear term 
 should be proportional to $\barD^\mu \barF^a_{\mu\nu}A^{a,\nu}$ (see the Appendix) which means that $J_0(\barA)$ is UV-finite.

 Looking at the diagrams in Fig. \ref{fig:bffig1} we see that $J_0(\Bara)$ differs from ${\delta \Gamma(\Bara)\over\delta\Bara}$ by replacement one of the bF vertices (\ref{vertexbF}) 
 with usual three-gluon vertex (\ref{vertex}). It is worth noting that in, say, scalar theory $J_0(\bfi)={\delta\Gamma(\bfi)\over\delta\bfi}$ and the complication in our case is due to the fact that
 $J_\mu$ in Eq. (\ref{tila}) depends both on $\Bara$ and $\tilA$.

 \section{Renormalization of twist-two gluon light-ray operator}
As an example, let us  consider the twist-two gluon LR  operator in pure gluodynamics:
\begin{eqnarray}
&&\hspace{-1mm}
\hat\calo_F ~=~g^2\hatF^a_{\xi n}(n)[n,0]^{ab}\hatF^{\xi,b}_{~n}(0)~
\label{gluLR}
\end{eqnarray}
where $n^2=0$. Here we use standard notations $n_\mu V^{\mu} \equiv V_n$ and  
$$
[x,y]~\equiv~{\rm Pexp}\Big\{ig\!\int_0^1\! du~(x-y)^\mu {A}_\mu(ux+\baru y)\Big\}
$$
for the straight-line ordered gauge link between points $x$ and $y$. Hereafter $\baru\equiv 1-u$.

As is well know, the counterterms in the Lagrangian (\ref{RenLag}) are not sufficient to regularize matrix elements of the operator
(\ref{gluLR}) so one needs to regularize this operator with extra counterterms. To find those, instead of considering matrix elements of 
our LR operator between gluon states, we  will consider the matrix element of the operator $\calo^F$ in the background field $\barA$ 
defined by Eq. (\ref{defiObF}).
\begin{eqnarray}
&&\hspace{-1mm}
\langle \hat\calo_F(A+\barA)\rangle ~=~\calo_F(\barA)
\label{fla28}\\
&&\hspace{-1mm}
+~g^2 f^{abc}\calF^\xi_{~n}(\lambda n)\langle \hatA^b_\xi(0) \hatA^c_n(0)\rangle_\barA 
+g^2 f^{abc}\langle \hatA^b_\xi(\lambda n) \hatA^c_n(\lambda n)\rangle_\barA \calF^\xi_{~n}(0)
\nonumber\\
&&\hspace{-1mm}
+~\langle (\barD_\xi \hatA^a_n-\barD_n\hatA^a_\xi)(\lambda n)(\barD^\xi \hatA^a_n-\barD_n\hatA^{\xi,a})(0)\rangle_\barA
\nonumber\\
&&\hspace{-1mm}
+~ig^2\langle (\barD_\xi \hatA^a_n-\barD_n\hatA^a_\xi)\!\int_0^\lambda \! du ([\lambda n,un]\hatA_n(un)[un,0])^{ab}\rangle_\barA
\calF^{\xi, b}_{~n}(0)
\nonumber\\
&&\hspace{-1mm}
+~ig^2\calF^{\xi, a}_{~n}(\lambda n)\langle \!\int_0^\lambda \! du ([\lambda n,un]\hatA_n(un)[un,0])^{ab}(\barD_\xi \hatA^b_n-\barD_n\hatA^b_\xi)\rangle_\barA+...
\nonumber
\end{eqnarray}
where $A^{ab}\equiv (T^m)^{ab}A^m$ in the adjoint representation ($(T^m)^{ab}=-if^{mab}$). The dots stand for higher-order terms in the expansion in quantum field $A$.  Note that in the RHS of Eq. (\ref{fla28}) 
we omitted terms $\barF^\alpha_{~n}(\lambda n)\langle (\barD_\alpha A^a_n-\barD_nA^a_\alpha\rangle_\barA$ 
and $\langle (\barD_\alpha A^a_n-\barD_nA^a_\alpha\rangle_\barA \barF^\alpha_{~n}(0)$ 
because they vanish due to Eq. (\ref{netua})

\begin{figure}[htb]
\includegraphics[width=88mm]{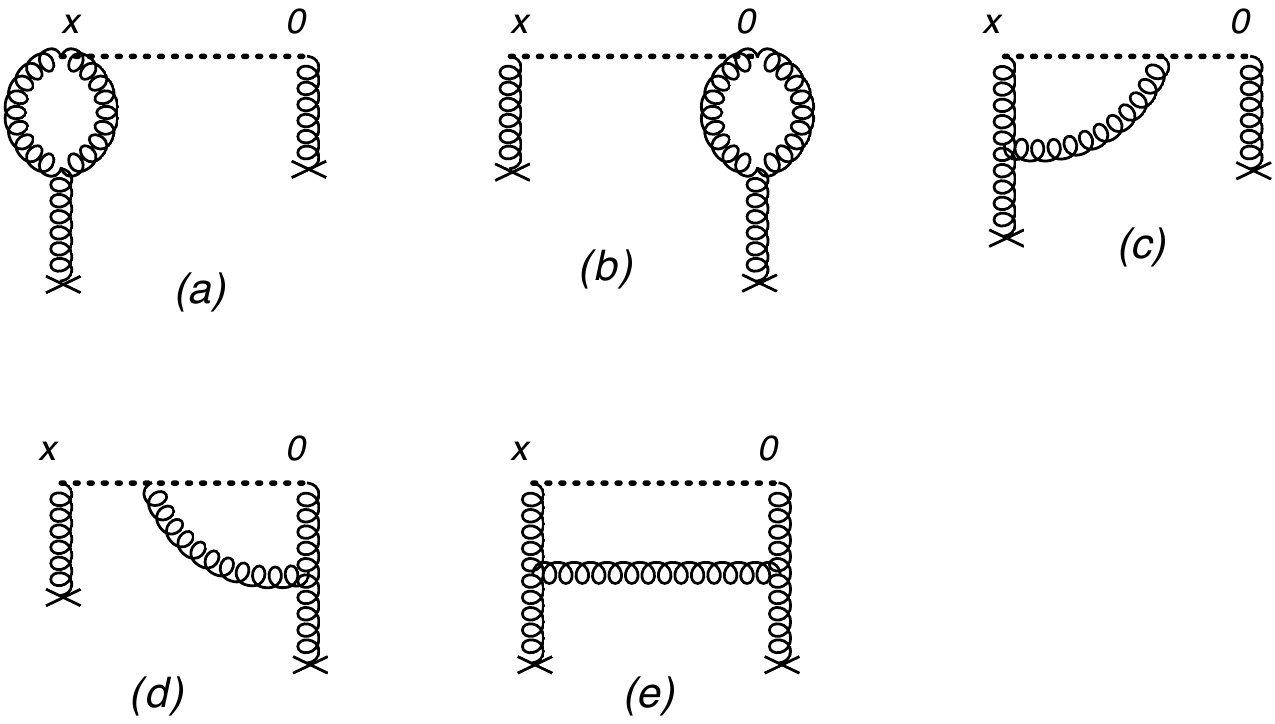}

\vspace{-22mm}
\caption{Gluon light-ray operator at one loop}
\label{fig:bffig2}
\end{figure}

The four terms in the RHS of this equation are
shown  in Fig. \ref{fig:bffig2}.
These diagrams were calculated in Ref.  \cite{Balitsky:1987k}
 and the result in has the form
\begin{eqnarray}
&&\hspace{-1mm}
\langle\hat\calo_F(A+\barA)\rangle~
\label{resultofcalc}\\
&&\hspace{-1mm}=~\calo_F(\barA)+{g^2\mu^{2\ve}\over 16\pi^2\ve}\!\int_0^1\!dudv~K(u,v)\calo_F(u,v;\barA)
~+~{\rm UV-finite~terms}
\nonumber
\ega
where 
\begin{eqnarray}
&&\hspace{-1mm}
\calo_F(un,vn;\barA) ~=~\balf^a_{\alpha n}(un)[\lambda n,0]^{ab}\balf^{\alpha}_{~n}(vn)~
\label{gluLRe}
\end{eqnarray}
and the gauge link is also made from $\barA$ fields. The gluon-gluon kernel has the form
\bega
&&\hspace{-1mm}
K(u,v)~=~-4(1-\baru -v+3\baru v)\theta(u-v)-\delta(\baru)\Big[{\barv^2\over v}+\delta(v)\!\int_0^1\!dv'{(\barv')^2\over v}\Big]
\nonumber\\
&&\hspace{-1mm}
-\delta(v)\Big[{u^2\over \baru}-\delta(\baru)\!\int_0^1\!du'{{u'}^2\over \baru}\Big]+3\delta(\baru)\delta(v)
\ega
where the convention $\int_0^1\! dx\delta(x)=1$ is assumed.

 The corresponding counterterm must subtract the UV divergence from Eq. (\ref{resultofcalc}) so
the renormalized LR operator $\hat\calo_F^\mu(n,\hatA)$ has the form
\begin{eqnarray}
&&\hspace{-1mm}
\hat\calo_F^\mu ~=~\hat\calo_F~
-{g^2(\mu)\over 16\pi^2\ve}\!\int_0^1\!dudv~K(u,v)\hat\calo_F(u,v)
\label{gluLRen}
\end{eqnarray}
where $\hat\calo_F(u,v)=g^2\hatF^a_{\xi n}(un)[un,vn]^{ab}\hatF^{\xi,b}_{~n}(vn)$ in accordance with Eqs. (\ref{gluLR}) 
and (\ref{gluLRe}).
Differentiating with respect to $\mu$ one obtains 
(recall that ${d\ln g(\mu)\over d\ln\mu}~=~-\ve-{g^2\over 16\pi^2}b_0$)
\beq
\mu{d\over d\mu}\hat\calo_F^\mu ~=~
-{g^2\mu^{2\ve}\over 16\pi^2\ve}\!\int_0^1\!dudv~K(u,v)\hat\calo_F^\mu(u,v)
\label{gluLRen}
\eeq
which corresponds to well-known result
\begin{eqnarray}
&&\hspace{-1mm}
\mu{d\over d\mu}[F^a_{\alpha n}(n)[n,0]^{ab}F^{\alpha}_{~n}(0)]^\mu
\\
&&\hspace{-1mm}
=~-{g^2N_c\over 4\pi^2}\!\int_0^1\!dudv~\Big[K(u,v)-{b_0\over 2N_c}\delta(\baru)\delta(v)\Big]
[F^a_{\alpha n}(un)[un,vn]^{ab}F^{\alpha}_{~n}(vn)]^\mu
\nonumber
\end{eqnarray}
%

\section{Light-cone expansion of one-loop propagator in a background field}
As was mentioned above, the Lagrangian (\ref{Lagrangian}) is relevant  for calculation of  diagrams 
with quantum fields beyond the tree approximation. For example, for the calculation of the one-loop 
gluon propagator in the background field one needs the counterterm 
\beq
\half \delta Z_3A^{a\mu}(\Bard^2 g_{\mu\nu}-2i\balf_{\mu\nu}-\barD_\mu\barD_\nu)^{ab}A^{b\nu}
\label{counterterm}
 \eeq
 to cancel the corresponding UV divergence in the loop. This was checked by explicit calculation of the 
 quark-loop contribution to gluon propagator in Ref.  \cite{Balitsky:2022vnb}.
 
 The quark-loop contribution to  gluon propagator in the bF (background-Feynman) gauge has the form
\begin{eqnarray}
&&\hspace{-1mm}
\langle \hatA^a_\mu(x) \hatA^b_\nu(y)\rangle_{\rm quark~loop}
~=~\int\!dz_1dz_2
(x|{1\over P^2g_{\mu\alpha}+2i\calf_{\mu\alpha}}|z_1)^{am}
\nonumber\\
&&\hspace{-0mm}
\times~ t^m\gamma_\alpha(z_1|{1\over\slashed{P}}|z_2)t^n\gamma_\beta(z_2|{1\over\slashed{P}}|z_1)
(z_2|{1\over P^2g_{\beta\nu}+2i\calf_{\beta\nu}}|y)^{nb}
\label{gluprop1}
\end{eqnarray}
To get the argument of coupling constant for the rapidity evolution of gluon TMD by BLM procedure \cite{Brodsky:1982gc}
 one  needs to calculate it near the light cone $(x-y)^2=0$ in the background field with the only component $\calf^{-i}(x^+)$
with one-$\calf$ accuracy.  The relevant diagrams for gluon propagator are shown in Fig. \ref{fig:bffig3}. 
\begin{figure}[htb]
\begin{center}
\includegraphics[width=111mm]{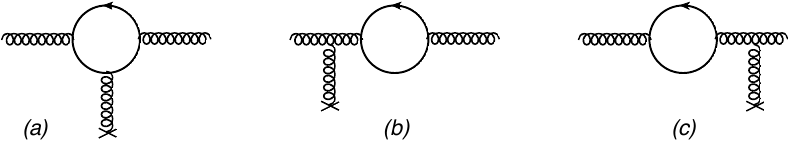}
\end{center}
\caption{Quark loop correction to gluon propagator in the background field.\label{fig:bffig3}}
\end{figure}
The UV parts of diagrams in Fig. \ref{fig:bffig3}b,c are canceled by 
$\half \delta Z_3^FA^{a\mu}(\partial^2g_{\mu\nu} -\partial_\mu\partial_\nu)^{ab}A^{b\nu}$ part of the counterterm  (\ref{counterterm})
which is present in the usual QCD Lagrangian (here $\delta Z_3^F=-{g^2n_f\over 24\pi^2\ve}$ is the quark part of $\delta Z_3$). 
On the contrary,  the UV divergence in Fig. \ref{fig:bffig3}a diagram requires full  Eq. (\ref{counterterm}) contribution, and 
the calculation of that diagram provides a check of the explicit form of counterterm (\ref{counterterm}). 

  After some simple but lengthly calculations one obtains
 \cite{Balitsky:2022vnb}
\begin{eqnarray}
&&\hspace{-2mm}
g^2{\rm Tr}~t^a\gamma_\alpha(z_1|{1\over\slP}|z_2)t^b\gamma_\beta(z_2|{1\over\slP}|z_1)-
{i\delta Z_3\over 2}(\Bard^2 g_{\mu\nu}-2i\balf_{\mu\nu}-\barD_\mu\barD_\nu)^{ab}\delta(z_{12})
\nonumber\\
&&\hspace{-2mm}
=~{ig^2\over 4\pi^2}\Big\{g_{\alpha\beta}(z_1|P^2\ln {\tmu^2\over -P^2}|z_2)
-(z_1|P_\alpha\ln{\tmu^2\over -P^2}P_\beta|z_2)
+ig\!\int_0^1\! du 
\nonumber\\
&&\hspace{11mm}
\times~\Big[u\Big(\calf_{\alpha \xi}(z_u)(z_1|{p^\xi\over p^2}|z_2)\Big)\stackrel{\leftarrow}P_\beta
-P_\alpha\Big(\baru \calf_{\beta \xi}(z_u)(z_1|{p^\xi\over p^2}|z_2)\Big)
\label{kvloop}\\
&&\hspace{-2mm}
+~(z_1|2\ln{\tmu^2\over -p^2}-{5\over 2}|z_2)\calf_{\alpha\beta}(z_u)
 +2i\baru u(z_1|{p^\xi\over p^2}|z_2)(D_\alpha \calf_{\beta \xi}(z_u)+\alpha\leftrightarrow\beta)\Big]
\Big\}
 \nonumber
\end{eqnarray}
where $\tmu^2\equiv \bar\mu^2_{\rm MS}e^{5/3}$. 
This form is very convenient for light-cone expansion since any function of $P^2$ can be rewritten in terms of integral of a heat kernel
(the light-cone expansion of heat kernels is presented in the Appendix).
Substituting this expression to Eq. (\ref{gluprop1}) and expanding near the light cone, we obtain after some algebra  \cite{Balitsky:2022vnb}
\begin{eqnarray}
&&\hspace{-2mm}
\langle \hatA^a_\mu(x) \hatA^b_\nu(y)\rangle_{\rm quark~loop}^{ab}~=~{g^2\over 24\pi^2}\bigg\{
i(x|g_{\mu\nu}{\ln {\tmu^2\over -p^2}\over p^2}-p_\mu p_\nu{\ln {\tmu^2\over -p^2}\over p^4}|y)
+~{i\over 8\pi^2\Delta^2}
\nonumber\\
&&\hspace{-2mm}
\times~\big[\ln{-\tmu^2 \Delta^2\over 4}-1+2\gamma \big]
\!\int_0^1\! du~[x,ux]\big(u\Delta_\nu \calf_{\mu\Delta}(x_u)-\baru \Delta_\mu \calf_{\nu\Delta}(x_u)\big)[ux,0]
\nonumber\\
&&\hspace{-2mm}
-~
{i\over 16\pi^2\Delta^2}\!\int_0^1\! du~ [x,ux]\big(\baru\ln\baru~
\Delta_\nu \calf_{\mu \Delta}(x_u)-u\ln u~ \Delta_\mu \calf_{\nu \Delta}(x_u))\big)[ux,0]
\nonumber\\
&&\hspace{-2mm}
+~{i\Gamma\big({d\over 2}-2\big)\over 32\pi^{d\over 2}(-\Delta^2)^{{d\over 2}-2}}
\!\int_0^1\! du~ [x,ux]\bigg(\Big[-{2\over d-4}-\ln{-\tmu^2 \Delta^2\over 4}
+~\psi\big({d\over 2}-1\big)-\gamma_E 
\nonumber\\
&&\hspace{33mm}
+~6-4\ln\baru u+u\ln u+\baru\ln\baru\Big]\calf_{\mu\nu}(x_u)
\label{gproponlico}\\
&&\hspace{-2mm}
+~\Big[{2\over d-4}+\ln{-\tmu^2 \Delta^2\over 4}-\psi\big({d\over 2}-1\big)-2+\gamma_E \Big]
\baru u [D_\mu \calf_{\nu \Delta}(x_u)+\mu\leftrightarrow\nu]
\nonumber\\
&&\hspace{-2mm}
-~ \big[u^2\ln uD_\mu \calf_{\nu \Delta}(x_u)
+\baru^2\ln\baru  D_\nu \calf_{\mu \Delta}(x_u)\big]\bigg)[ux,0]\bigg\}^{ab}
~+~O\big(D^\mu \calf_{\mu \nu}, \calf^2\big)
\nonumber
\end{eqnarray}
where $\Delta\equiv x-y$, $x_u\equiv ux+\baru y$, $\psi$ is the logarithmic derivative of gamma-function, and $\gamma_E$ is the Euler constant. 
The gluon loop contribution to gluon propagator in the background field can be obtained in a similar way, although the calculations are expected to be much more involved.

\section{Conclusions}

As outlined in the Introduction, the background-field method is commonly employed to derive a factorization for certain processes by applying successive integration over fields in the functional integral. In this framework, fields with momenta above a relevant cutoff (such as transverse momentum in collinear factorization or longitudinal momentum in rapidity factorization) are treated as quantum fields, while those with momenta below the cutoff are treated as background fields. A fundamental requirement is that quantum field cannot turn into background field(s).

At the one-loop level, this distinction is clear and unambiguous. However, at higher loops, this requirement becomes less well-defined. In this context, the Lagrangian (\ref{Lagrangian}) formalizes the condition that quantum fields cannot be transformed into ``classical'' background fields.

As previously noted, calculating the effective action does not require the renormalization of quantum fields or the inclusion of counterterms for both background and quantum fields in the Lagrangian. However, to extend beyond the effective action and, for instance, compute one-loop propagators in background fields near the light cone, it is necessary to account for the full set of counterterms as outlined in Eq. (\ref{Lagrangian_bF}).

\section*{Acknowlwdgements}
The author is grateful  to V. Braun and A. Vladimirov for valuable discussions. 
This work  is supported by DOE contract DE-AC05-06OR23177  and by the grant DE-FG02-97ER41028.
This work is also supported by the U.S. Department of Energy, Office of Science, Office of Nuclear Physics, within the framework of the Saturated Glue (SURGE) Topical Theory Collaboration.

\section*{Appendix}
\subsection*{The source $J_0(\barA)$ at  one-loop level in all orders in the background field}
In this Section we calculate the explicit form of the source $\delta J_0(\Bara)$ in the leading order in $g^2$ but in all orders in the
background field. The  quantum field in the $\Bara$ background is given by diagrams in Fig. \ref{fig:bffig3}
\begin{figure}[htb]
\includegraphics[width=111mm]{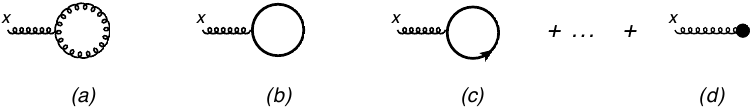}
\caption{Quantum field in $\barA$ background. The lines are propagators in the background field given by Eq. (\ref{props})}
\label{fig:bffig4}
\end{figure}
and the source $\delta J_0$ should be chosen such that $\langle \hatA\rangle_\Bara=0$. 

Let us first calculate contribution of the diagram in Fig. \ref{fig:bffig4}a.  We get
\begin{eqnarray}
&&\hspace{-1mm}
 \langle \hatA_\alpha(x)\rangle_\barA~
 \stackrel{\rm Fig. ~\ref{fig:bffig4}a}=~ \langle \hatA^a_\alpha(x)\exp\big\{-ig\mu^\ve\!\int dz~
f^{mnl}\hatA^m_{\mu}\hatA^n_{\nu}(D^{\mu}\hatA^{\nu})^l\big\}\rangle_\barA
\nonumber\\
&&\hspace{-1mm}
=~-ig\mu^\ve f^{mnl}\!\int\! dz\langle \hatA^a_\alpha(x) \hatA^m_{\mu}(z)\rangle\langle \hatA^n_{\nu}(z)(D^{\mu}\hatA^{\nu}(z))^l-\mu\leftrightarrow\nu\rangle 
\nonumber\\
&&\hspace{-1mm}
-~{i\over 2}gf^{mnl}\!\int\! dz\langle  \hatA^a_\alpha(x) (D^{\mu}\hatA^{\nu}(z))^l-\mu\leftrightarrow\nu\rangle\langle \hatA^m_{\mu}(z)\hatA^n_{\nu}(z)\rangle 
\nonumber\\
&&\hspace{-1mm}
=~ig\mu^\ve f^{mnl}\!\int\! dz\langle \hatA^a_\alpha(x) \hatA^{m,\mu}(z)\rangle_\Bara\Big(\langle\barD^\mu \hatA^n_\nu(z)\hatA^l_\nu(z)\rangle_\Bara
\nonumber\\
&&\hspace{-1mm}
-2\langle\barD^\nu \hatA^n_\mu(z)\hatA^l_\nu(z)\rangle_\Bara+\langle\barD^\nu \hatA^n_\nu(z)\hatA^l_\mu(z)\rangle_\Bara\Big)
\nonumber\\
&&\hspace{-1mm}
=~ig\mu^\ve f^{mnl}\!\int\! dz\langle \hatA^a_\alpha(x) \hatA^{m,\mu}(z)\rangle_\Bara(z|-P_\mu \big({1\over P^2+2i\balF }\big)_{\nu\nu}
\nonumber\\
&&\hspace{-1mm}
+~2P_\nu \big({1\over P^2+2i\balF }\big)_{\mu\nu}-P_\nu \big({1\over P^2+2i\balF }\big)_{\nu\mu}|z)^{nl}
\label{fla39}
\end{eqnarray}
Let us find the UV-divergent part of this contribution. 
Using formulas (\ref{localcounter1}) and  (\ref{localcounter2}) one easily obtains
\begin{eqnarray}
&&\hspace{-1mm}
 \langle \hatA^a_\alpha(x)\rangle_\barA^{\rm UV}~\stackrel{\rm Fig. ~\ref{fig:bffig4}a}=~-{ig^2N_c\over 16\pi^2}{5\over 2\ve}\!\int\! dz
\langle  \hatA^a_\alpha(x) \hatA^{m,\mu}(z)\rangle \barD^\xi \barF^m_{\xi\mu}(z)
 \label{glUV}
\end{eqnarray}
in accordance with Eq. (\ref{pertvkladglu}).
Note that due to explicit gauge invariance of Eq.  (\ref{localcounter1}) we obtain the above result in the gauge-invariant form.
In particular,  this means that the UV parts of diagrams in Figs. \ref{fig:bffig1}d and \ref{fig:bffig1}d are given by the non-Abelian
terms in $\barD_\xi \balf^{\xi\mu}(z)$.

Similarly, one gets for the ghost diagram in Fig. \ref{fig:bffig4}b
\begin{eqnarray}
&&\hspace{-1mm}
\langle\hatA^a_\alpha(x)\rangle_\barA~\stackrel{\rm Fig. ~\ref{fig:bffig4}b}=~ g\mu^\ve\!\int\! dz~ \langle \hatA^a_\alpha(x)\barc^m\stackrel{\leftarrow}{\Bard^\mu} \hatA_\mu^{mn}c^n(z)\rangle_\barA
\nonumber\\
&&\hspace{-1mm}
=~ ig\mu^\ve f^{mnl}\!\int\! dz~ \langle \hatA^a_\alpha(x)\hatA^m_\mu(z)\rangle(z|P^\mu{1\over P^2}|z)^{nl}
\end{eqnarray}
and the UV part according to Eq. (\ref{localcounter2}) has the form
\begin{eqnarray}
&&\hspace{-1mm}
 \langle \hatA^a_\alpha(x)\rangle_\barA^{\rm UV}~\stackrel{\rm Fig. ~\ref{fig:bffig4}b}
 =~-{ig^2N_c\over 16\pi^2}{1\over 6\ve}\!\int\! dz\langle  \hatA^a_\alpha(x) \hatA^{m,\mu}(z)\rangle \barD^\xi \barF^m_{\xi\mu}(z)
 \label{ghUV}
\end{eqnarray}
For the quark contribution in diagram Fig. ~\ref{fig:bffig4}c one obtains
\begin{eqnarray}
&&\hspace{-1mm}
\langle\hatA^a_\alpha(x)\rangle_\barA~\stackrel{\rm Fig. ~\ref{fig:bffig4}c}=~ 
ig\mu^\ve\!\int\! dz~ \langle \hatA^a_\alpha(x)\hbsi \hatA_\mu\gamma^\mu\hsi(z)\rangle_\barA\\
&&\hspace{-1mm}
=~g\mu^\ve\!\int\! dz~ \langle \hatA^a_\alpha(x) \hatA_\mu^m(z)\rangle {\rm Tr}\{t^m\gamma^\mu (z|\slP{1\over P^2+\half\sigma\balf}|z)
\nonumber
\end{eqnarray}
and the UV-divergent part is
\begin{eqnarray}
&&\hspace{-1mm}
 \langle \hatA^a_\alpha(x)\rangle_\barA^{\rm UV}~\stackrel{\rm Fig. ~\ref{fig:bffig4}c}
 =~{ig^2\over 16\pi^2}{2\over 3\ve}\!\int\! dz\langle  \hatA^a_\alpha(x) \hatA^{m,\mu}(z)\rangle \barD^\xi \barF^m_{\xi\mu}(z)
 \label{kvUV}
\end{eqnarray}
due  to Eq. (\ref{localcounter3}). 

Adding these contributions we obtain
\begin{eqnarray}
&&\hspace{-1mm}
 \langle A_\alpha(x)\rangle_\barA~\stackrel{\rm Fig. ~\ref{fig:bffig4}a-c}=~
 \\
&&\hspace{-1mm}
=~g\mu^\ve\!\int\! dz\langle A^a_\alpha(x) A^{b,\mu}(z)\rangle_\Bara
\Big[if^{bcd}(z|-P_\mu \big({1\over P^2+2i\balF }\big)_{\nu\nu}
\nonumber\\
&&\hspace{-1mm}
+~2P_\nu \big({1\over P^2+2i\balF }\big)_{\mu\nu}-P_\nu \big({1\over P^2+2i\balF }\big)_{\nu\mu}+P^\mu{1\over P^2}|z)^{cd}
\nonumber\\
&&\hspace{-1mm}
+~n_f {\rm Tr}\{t^b\gamma^\mu (z|\slP{1\over P^2+\half\sigma\balf}|z)\Big]
\nonumber
\end{eqnarray}
and the UV-divergent part is
\begin{eqnarray}
&&\hspace{-1mm}
 \langle \hatA_\alpha(x)\rangle_\barA^{\rm UV}~\stackrel{\rm Fig. ~\ref{fig:bffig3}a-c}=~-{ig^2\over 16\pi^2}\Big[{8\over 3}N_c-{2\over 3\ve}n_f\Big]\!\int\! dz\langle  \hatA^a_\alpha(x) \hatA^{m,\mu}(z)\rangle \barD^\xi \barF^m_{\xi\mu}(z)
 \nonumber\\
\end{eqnarray}
Finally, we need to add contribution of the last term n Eq. (\ref{Lagrangian_bF}) schematically shown in Fig. ~\ref{fig:bffig4}d.
\beq
 \langle \hatA_\alpha(x)\rangle_\barA~\stackrel{\rm Fig. ~\ref{fig:bffig4}d}
=~i\!\int\! dz\langle \hatA^a_\alpha(x) \hatA^{b,\mu}(z)\rangle_\Bara\big[\half(\delta Z+\delta Z_3)\barD_\mu \barF^{b,\mu\nu}-\delta J_0^{b,\nu}\big]
\eeq
From Eq. (\ref{Zisoneloop}) we see that $\half(\delta Z+\delta Z_3)~=~{g^2\over 16\pi^2\ve}\big({8\over 3}N_c-{2\over 3}n_f\big)$ 
so the UV part of the contribution of diagrams Fig. ~\ref{fig:bffig3}a-c  is canceled by the contribution of the counterterm $\half(\delta Z+\delta Z_3)$. The remaining final part should be canceled by $\delta J_0$ so we get

\bega
&&\hspace{-1mm}
\delta J_0^{b\nu}(z)~=~g\mu^\ve f^{bcd}(z|-P_\mu \big({1\over P^2+2i\balF }\big)_{\nu\nu}
\nonumber\\
&&\hspace{-1mm}
+~2P_\nu \big({1\over P^2+2i\balF }\big)_{\mu\nu}-P_\nu \big({1\over P^2+2i\balF }\big)_{\nu\mu}+P^\mu{1\over P^2}|z)^{cd}
\nonumber\\
&&\hspace{-1mm}
-~ig\mu^\ve n_f {\rm Tr}\{t^b\gamma^\mu (z|\slP{1\over P^2+\half\sigma\balf}|z)~-~{\rm UV~pole}
\label{deltage}
\ega
Note that the first term of the expansion of the RHS in powers of $\barA$ agrees with Eq. (\ref{deltage1}). 
However, the full expression (\ref{deltage}) is gauge invariant.

\subsection*{Heat kernel expansions}
To obtain the expansion of propagators (\ref{props}) near $x=0$ it is convenient to use
 the representation  in terms of the integrals of corresponding  ``heat kernels''. 
 Let us start with a scalar propagator
\beq
(x|{1\over P^2+i\epsilon}|y)~=~-i\!\int_0^\infty \! ds ~(x|e^{isP^2}|0)~=~\!\int_0^\infty \! ds ~(x|e^{is(p^2+\{p,\bala\}+\bala^2)}|0)
\label{scalprop}
\eeq
Expanding the  $e^{is(\{p,\bala\}+\bala^2)}$ in powers of the proper time $s$ and using formulas from  Ref. \cite{Balitsky:1987k} 
one obtains
\begin{eqnarray}
&&\hspace{-1mm}
(x|e^{is(P^2-m^2)}|0)~
\label{scalheat}\\
&&\hspace{-1mm}
=~(x|e^{is(p^2-m^2)}|0)\Big\{[x,0]+s\!\int_0^1\! du~\baru u[x,ux]\barD^\mu \balF_{\mu\nu}x^\nu(ux)[ux,0]
\nonumber\\
&&\hspace{-1mm}+2is\!\int_0^1\! du\!\int_0^u\!dv~\baru v[x,ux]x_\mu \balF^{\mu\xi}(ux)[ux,vx]x^\nu \balF_{\nu\xi}(vx)[vx,0]
\nonumber\\
&&\hspace{-1mm}
+~
2s^2\!\int_0^1\! du\!\int_0^u\! dv~[x,ux]\big(\baru v\balF_{\xi\eta}(u)[ux,vx]\balF^{\xi\eta}(v)+\baru^2v^2
\nonumber\\
&&\hspace{-1mm}
\times~
x_\lambda x^\rho D_\eta F^{\lambda\xi}(u)[ux,vx]\barD^\eta \balF_{\rho\xi}(v)\big)[vx,0]\Big\}
~+~O(\barD^\xi \balF_{\xi\eta} \balF_{\mu\nu},\balF^3)
\nonumber
\end{eqnarray}
We also need a heat kernel for gluon operator
\begin{eqnarray}
&&\hspace{-1mm}
(x|e^{is(P^2+2i\balF-m^2)}|0)_{\alpha\beta}~
\nonumber\\
&&\hspace{-1mm}
=~(x|e^{is(p^2-m^2)}|0)
~\Big\{[x,0]+sg_{\alpha\beta}\!\int_0^1\! du~\baru u[x,ux]\barD^\mu \balF_{\mu x}(ux)[ux,0]
\nonumber\\
&&\hspace{-1mm}
+~2isg_{\alpha\beta}\!\int_0^1\! du\!\int_0^u\!dv~\baru v[x,ux]\balF_x^{~\xi}(ux)[ux,vx]\balF_{x\xi}(vx)[vx,0]
\nonumber\\
&&\hspace{-1mm}
-~2s\! \int_0^1\! du ~[x,ux]\balF_{\alpha\beta}(ux)[ux,0]
\nonumber\\
&&\hspace{-1mm}
+~s^2\!\int_0^1\! du~[x,ux]\Big\{2i\baru u \barD^2\balF_{\alpha\beta}(ux)[ux,0]
\nonumber\\
&&\hspace{-1mm}
+~\!\int_0^u\! dv\Big[4\balF_{\alpha\xi}(ux)[ux,vx]\balF^\xi_{~\beta}(vx)
-4\baru v\big(\barD^\xi \balF_{\alpha\beta}(ux)[ux,vx]\balF_{\xi x}(vx)
\nonumber\\
&&\hspace{-1mm}
+~\balF_{\xi x}(ux)[ux,vx]\barD^\xi \balF_{\alpha\beta}(vx)\big)
+
2g_{\alpha\beta}\big(\baru v\balF_{\xi\eta}(u)[ux,vx]\balF^{\xi\eta}(v)
\nonumber\\
&&\hspace{-1mm}
+\baru^2v^2 \barD_\eta \balF_x^{~\xi}(u)[ux,vx]\barD^\eta \balF_{x\xi}(v)\big)\Big][vx,0]\Big\}
\nonumber\\
&&\hspace{-1mm}
-~4is^3\!\int_0^1\! du\!\int_0^u\! dv[x,ux]\Big[\baru^2v^2\big(\barD^\eta \barD^\xi \balF_{\alpha\beta}(ux)[ux,vx]\barD_\eta \balF_{x\xi}(vx)
\nonumber\\
&&\hspace{-1mm}
+\barD_\eta \balF_{x\xi}(ux)[ux,vx]\barD^\eta \barD^\xi \balF_{\alpha\beta}(vx)\big)
\label{gluonheat}\\
&&\hspace{-1mm}
-~2\baru v \barD^\lambda \balF_{\alpha\xi}(ux)[ux,vx]\barD_\lambda \balF_\beta^{~\xi}(vx)\Big][vx,0]\Big\}
~+~O(\barD^\xi \balF_{\xi\eta} \balF_{\mu\nu},\balF^3)
\nonumber
\end{eqnarray}
so the gluon propagator (with IR regulator $m$) has the form
\begin{eqnarray}
&&\hspace{-1mm}
(x|{1\over P^2+2i\balF-m^2}|0)_{\alpha\beta}~=~(x|{1\over p^2-m^2}|0)[x,0]
\nonumber\\
&&\hspace{-1mm}
+~(x|{i\over (p^2-m^2)^2}|0)\Big\{\!\int_0^1\! du~[x,ux]\Big(g_{\alpha\beta}\baru u\barD^\mu \balF_{\mu x}(ux)[ux,0]
\nonumber\\
&&\hspace{-1mm}
-~2\balF_{\alpha\beta}(ux)[ux,0]+2ig_{\alpha\beta}\!\int_0^u\!dv~\baru v \balF_x^{~\xi}(ux)[ux,vx]\balF_{x\xi}(vx)[vx,0]\Big)\Big\}
\nonumber\\
&&\hspace{-1mm}
-~(x|{2\over (p^2-m^2)^3}|0)\!\int_0^1\! du~[x,ux]\Big\{2i\baru u \barD^2\balF_{\alpha\beta}(ux)[ux,0]
\nonumber\\
&&\hspace{-1mm}
+~\!\int_0^u\! dv\Big[4\balF_{\alpha\xi}(ux)[ux,vx]\balF^\xi_{~\beta}(vx)
-4\baru vx^\eta\Big(\barD^\xi \balF_{\alpha\beta}(ux)[ux,vx]\balF_{\xi\eta}(vx)
\nonumber\\
&&\hspace{-1mm}+\balF_{\xi\eta}(ux)[ux,vx]\barD^\xi \balF_{\alpha\beta}(vx)]\Big)
+~
2g_{\alpha\beta}\Big(\baru v\balF_{\xi\eta}(u)[ux,vx]\balF^{\xi\eta}(v)
\nonumber\\
&&\hspace{-1mm}
+~\baru^2v^2 \barD_\eta \balF_x^{~\xi}(u)[ux,vx]\barD^\eta \balF_{x\xi}(v)\Big)\Big][vx,0]\Big\}~+~O\Big((x|{1\over (p^2-m^2)^4}|0)\Big)   
\nonumber\\
&&\hspace{-1mm}
\end{eqnarray}
For the calculation of contribution of UV-divergent  parts of Fig. 3  diagrams we need a UV part of  
$(0|P_\mu{1\over P^2g_{\alpha\beta}+2i\balF_{\alpha\beta}-m^2}|0)$. 
Using formula
\begin{eqnarray}
&&\hspace{-1mm}
{\partial\over\partial x_\mu}[ux,vx]~
\\
&&\hspace{-1mm}
=~iu\bala_\mu(ux)[ux,vx]-[ux,vx]iv\bala_\mu(vx)-i\!\int_v^u\! dt ~t[ux,tx]x^\rho \balF_{\rho\mu}(tx)[tx,vx]
\nonumber
\end{eqnarray}
one quickly realizes that the UV part of $\lim_{x\rightarrow 0}(x|P_\mu{1\over P^2+2i\balF-m^2}|0)_{\alpha\beta}$ can come only from
\begin{eqnarray}
&&\hspace{-1mm}
(0|{i\over (p^2-m^2)^2}|0)\lim_{x\rightarrow 0}\big(i{\partial\over\partial x^\mu}+\bala^\mu)\Big\{\!\int_0^1\! du~[x,ux]\Big(g_{\alpha\beta}\baru u\barD^\xi \balF_{\xi x}(ux)[ux,0]
\nonumber\\
&&\hspace{-1mm}
-~2\balF_{\alpha\beta}(ux)[ux,0]
+~2ig_{\alpha\beta}\!\int_0^u\!dv~\baru v \balF_x^{~\xi}(ux)[ux,vx]\balF_{x\xi}(vx)[vx,0]\Big)\Big\}
\nonumber\\
&&\hspace{-1mm}
=~(0|{1\over (p^2-m^2)^2}|0)\big[\barD_\mu \balf_{\alpha\beta}(0)-{g_{\alpha\beta}\over 6}\barD^\xi\balf_{\xi\mu}(0)\big]
\end{eqnarray}
Thus,
\begin{eqnarray}
&&\hspace{-1mm}
\lim_{x\rightarrow 0}(x|P_\mu{1\over P^2+2i\balF-m^2}|0)_{\alpha\beta}~
=~{i\over 16\pi^2\ve}\big[\barD_\mu \balF_{\alpha\beta}(0)-{g_{\alpha\beta}\over 6}\barD^\xi\balF_{\xi\mu}(0)\big]~+~...
\nonumber\\
\label{localcounter1}
\end{eqnarray}
Looking at terms $\sim g_{\alpha\beta}$ one quickly gets the UV-divergent part of 
$\lim_{x\rightarrow 0}(x|{1\over P^2-m^2}|0)$
\begin{eqnarray}
&&\hspace{-1mm}
\lim_{x\rightarrow 0}(x|P_\mu{1\over P^2-m^2}|0)^{ab}~=~-\lim_{x\rightarrow 0}(x|{1\over P^2-m^2}P_\mu|0)^{ba}
\nonumber\\
&&\hspace{-1mm}
~=~-{i\over 16\pi^2\ve}{1\over 6}\barD^\xi\balF_{\xi\mu}^{ab}(0)~+~...
\label{localcounter2}
\end{eqnarray}
For quark contribution, we need also
\begin{eqnarray}
&&\hspace{-1mm}
\lim_{x\rightarrow 0}(x|P_\mu{1\over P^2+{i\over 2}\balf_{\alpha\beta}-m^2}|0)~
=~
~{1\over 16\pi^2\ve}\big[{1\over 4}\barD_\mu \sigma\balF(0)-{i\over 6}\barD^\xi\balF_{\xi\mu}(0)\big]~+~...
\nonumber\\
\label{localcounter3}
\end{eqnarray}
The structure  in the LHS is the same as in Eq. (\ref{localcounter1}) so we just replaced $2i\balf_{\alpha\beta}$ by $\half\sigma \balf$.
Multiplying by $\gamma^\mu$ we get the UV part of $O(g^2)$ quark contribution in the form
\begin{eqnarray}
&&\hspace{-1mm}
\lim_{x\rightarrow 0}(x|\slP{1\over P^2+{1\over 2}\balf_{\alpha\beta}-m^2}|0)~=~
~{i\over 16\pi^2\ve}{1\over 3}\barD^\xi\balF_{\xi\eta}(0)\gamma_\eta~+~{\rm UV-finite~terms}
\nonumber\\
\label{localcounter3}
\end{eqnarray}
For many flavors of massless quarks this expression should be multiplied by $n_f$.


\bibliography{bFact.bib}

\end{document}